\documentclass[letterpaper, 10 pt , conference]{ieeeconf}

\IEEEoverridecommandlockouts                              

\usepackage{graphicx} 
\usepackage{epsfig}
\usepackage{epstopdf}
\usepackage{stmaryrd}
\usepackage{cite} 
\usepackage{amsmath} 
\usepackage{amssymb}  
\usepackage{mathrsfs}
\usepackage{color}
\usepackage{nicefrac}
\usepackage[linesnumbered, ruled, vlined]{algorithm2e}
\usepackage{mathtools}
\mathtoolsset{showonlyrefs}

\usepackage{amsthm}

\usepackage{wrapfig,graphicx}
\usepackage{hyperref}
\graphicspath{{.}}

\theoremstyle{plain}

\newtheorem{problem}{Problem}
\usepackage{etoolbox}
\usepackage{float}
\usepackage{tabularx}

\def\th@definition{
	\thm@headfont{\itshape} 
	\thm@notefont{} 
}
\makeatother

\usepackage{color}

\theoremstyle{definition}
\newtheorem{definition}{Definition}

\newtheorem{lemma}{Lemma}
\newtheorem{example}{Example}
\newtheorem{theorem}{Theorem}
\newtheorem{remark}{Remark}

\newcommand{\adj}{\textnormal{Adj}}

\makeatletter
\patchcmd{\@makecaption}
{\scshape}
{}
{}
{}
\makeatother

\makeatletter
\def\endthebibliography{%
	\def\@noitemerr{\@latex@warning{Empty `thebibliography' environment}}%
	\endlist
}
\makeatother

\title{\LARGE \bf
Edge Differential Privacy for Algebraic Connectivity of Graphs
}

\author{Bo Chen, Calvin Hawkins, Kasra Yazdani, Matthew Hale$^{\ast}$
\thanks{$^{\ast}$Department of  Mechanical and Aerospace Engineering at the University of Florida, Gainesville, FL USA. Emails: \texttt{\{bo.chen,calvin.hawkins,kasra.yazdani,matthewhale\}} \texttt{@ufl.edu}. This work was supported in part by NSF under CAREER Grant~{\#}1943275 and by AFOSR under Grant~{\#}FA9550-19-1-0169. 
}
}

\begin{document}

\maketitle

\begin{abstract}
Graphs are the dominant formalism for modeling multi-agent 
systems. 
The algebraic connectivity of a graph is particularly important
because it provides the convergence rates of consensus algorithms
that underlie many multi-agent control and optimization techniques.
However, sharing the value of algebraic connectivity can inadvertently
reveal sensitive information about the topology of a graph, such as
connections in social networks. 
Therefore, in this work we present a method to release a graph's
algebraic connectivity under 
a graph-theoretic form of differential privacy, called
\emph{edge differential privacy}. Edge differential privacy obfuscates differences
among graphs' edge sets and thus conceals the absence or presence
of sensitive connections therein. 
We provide privacy with bounded Laplace noise,
which improves accuracy relative to conventional unbounded noise.
The private algebraic connectivity values are analytically
shown to provide accurate estimates of consensus convergence rates,
as well as accurate bounds on the diameter of a graph and the mean
distance between its nodes. Simulation results confirm the utility
of private algebraic connectivity in these contexts. 
\end{abstract}
\section{Introduction}
Graphs are used to model a wide range
of interconnected systems, including multi-agent control
systems~\cite{Ren2005}, social networks~\cite{Scott1988}, and others~\cite{SHIRLEY2005287}.
Various properties of these graphs have been used to
analyze controllers and dynamical processes over them,
such as reaching a consensus~\cite{zheng_consensus_2011}, 
the spread of a virus~\cite{Mieghem2009}, robustness to 
connection failures~\cite{freitas2020evaluating}, and others.
Graphs in these applications may contain sensitive information,
e.g., one's close friendships in the case of a social network,
and it is essential that these analyses do not inadvertently
leak any such information. 

Unfortunately, it is well-established that even graph-level analysis
may inadvertently reveal sensitive information about graphs,
such as the absence or presence of individual nodes in a graph~\cite{kasiviswanathan13}
and the absence or presence of specific edges between them~\cite{karwa14}.
This privacy threat has received attention in the
data science community, where graphs represent datasets
and the goal is to enable data analysis while safeguarding
the data of individuals in those datasets. 

Differential privacy is one well-studied tool for doing so.
Differential privacy is a statistical notion of privacy that has several
desirable properties: (i) it is robust to side information, in that learning
additional information about data-producing entities does not weaken privacy
by much~\cite{kasiviswanathan14}, and (ii) it is immune to post-processing, in that arbitrary
post-hoc computations on private data do not weaken privacy~\cite{dwork_algorithmic_2013}. 
There exist numerous differential privacy implementations for
graph properties, including counts
of subgraphs~\cite{karwa14}, 
degree distributions~\cite{day16}, and 
other frequent patterns in graphs~\cite{shen13}. 
These privacy mechanisms generally follow
the pattern of computing the quantity of interest, adding
carefully calibrated noise to it, and releasing its
noisy form. Although simple, this approach strongly
protects data with a suite of guarantees provided
by differential privacy~\cite{dwork_algorithmic_2013}. 

In this paper, we develop an edge differential privacy
mechanism to protect the algebraic connectivity of graphs.
A graph's algebraic connectivity (also called its Fiedler value)
is equal to the second-smallest eigenvalue of its Laplacian.
This value plays a central role in the study of multi-agent
systems because it sets the convergence rates
of consensus algorithms~\cite{olfati04}, which appear directly or in modified
form in formation control~\cite{ren07}, 
connectivity control~\cite{gennaro06}, 
and many distributed optimization 
algorithms~\cite{nedic18}.

As with existing graph analyses,
even the scalar-valued algebraic connectivity
 poses a significant privacy threat. 
We illustrate this point with two concrete privacy attacks
that enable inferences about the presence of certain edges in a graph.
These edges can represent, e.g., social connections between
individuals, and 
such applications require privacy protections 
when a graph's algebraic connectivity is shared. 

We therefore protect a graph's algebraic connectivity using
edge differential privacy, which 
obfuscates the absence and/or presence of a pre-specified number
of edges. Our implementation uses the recent bounded Laplace
mechanism~\cite{holohan2018bounded}, which ensures that private scalars lie in
a specified interval. The algebraic connectivity of a graph
is bounded below by zero and above by the number of nodes
in a graph, and we confine private outputs
to this interval. 

We provide closed-form values for the sensitivity
and other constants needed to define a privacy mechanism
for algebraic connectivity, and this is the first contribution
of this paper. The second contribution is bounding the error
that privacy induces in the convergence rates of consensus.
Differential privacy has made inroads in control
applications ranging from LQ control~\cite{yazdani2018differentially}, state
estimation~\cite{yazdani2020error}, formation control~\cite{hawkins2020differentially}, Markov decision
processes~\cite{gohari2020privacy} and others, due in part to the high performance
one can maintain even with privacy implemented. We show
that this is the case for the consensus setting as well.
Our third contribution is the use of the private values
of algebraic connectivity to bound other graph properties, namely
the diameter of graphs and the mean distance between their nodes. 

We note that~\cite{wang13} has developed a different
approach to privacy for the eigendecomposition of
a graph's adjacency matrix. Given our motivation
by multi-agent systems, we focus
on a graph's Laplacian, and we derive
simpler forms for the distribution of noise required,
as well as a privacy mechanism that does not
require any post-processing. 

The rest of the paper is organized as follows.
Section~II provides 
background, examples of privacy attacks,
and problem statements.
Section~III develops the differential privacy
mechanism for algebraic connectivity.
Next, Section~IV uses this mechanism to privately
compute consensus convergence rates,
and Section~V applies it to bounding other
graph properties. Then, Section~VI provides
simulation results and Section~VII provides
concluding remarks.

\section{Background and Problem Formulation}\label{sec:problemFormulation}
In this section, we briefly review background on graph theory and differential privacy, followed by formal problem statements. 

\subsection{Background on Graph Theory}
We consider an undirected, unweighted graph~$G = (V,E)$ defined over a set of nodes $V = \{1,\dots, n \}$
with edge set $E\subset V\times V$. The pair $(i,j)$ belongs to $E$ if nodes $i$ and $j$ share an edge, and $(i,j) \notin E$ otherwise. 
    Let~$\mathcal{G}_n$ denote the set of all graphs on~$n$ nodes. 
We let $d_i = \vert \{ j\in V \mid (i,j) \in E\}\vert$ denote the degree of node $i\in V$. The degree matrix $D(G)\in \mathbb{R}^{n\times n}$ is the diagonal matrix $D(G)=\operatorname{diag}\big(d_{1}, \ldots, d_{n}\big)$. The adjacency of~$G$ is
\begin{equation}
    (H(G))_{i j}=\begin{cases}
1 & (i, j) \in E \\
0 & \text { otherwise }
\end{cases}.
\end{equation}
We denote the Laplacian of graph $G$ by~${L(G)=D(G)-H(G)}$, which we simply refer to by $L$ when the associated graph is clear from  context. 

Let the eigenvalues of~$L$ be ordered
according to~$\lambda_1(L) \leq \lambda_2(L) \leq \cdots \leq \lambda_n(L)$. 
The matrix~$L$ is symmetric and positive semidefinite, and thus~$\lambda_i(L) \geq 0$ for all~$i$. 
All graphs~$G$ have~$\lambda_1(L) = 0$, and a seminal result shows that~$\lambda_2(L) > 0$
if and only if~$G$ is connected \cite{fiedler1975property}. 
Thus,~$\lambda_2$ is often called the \emph{algebraic connectivity} of a graph. Throughout this paper,
we consider connected graphs. 

The value of~$\lambda_2$ encodes a great deal of
information about~$G$: its value is non-decreasing in the number
of edges in~$G$, and algebraic connectivity is closely related to graph diameter and various other algebraic
properties of graphs~\cite{abreu07}. The value of~$\lambda_2$ also characterizes the performance
of consensus algorithms. Specifically, disagreement in a consensus protocol decays proportionally to~$e^{-\lambda_2 t}$ \cite{Mesbahi2010}.


\subsection{Background on Differential Privacy}
Differential privacy is enforced by a \emph{mechanism}, which is a randomized map. 
Given ``similar" inputs, a differential privacy mechanism produces outputs that are approximately indistinguishable from each other. 
Formally, a mechanism must obfuscate differences between inputs that are 
\emph{adjacent}\footnote{The word ``adjacency'' appears in two forms in this paper: for the adjacency matrix~$H$ above,
and for the adjacency relation used by differential privacy. The adjacency matrix appears only in this section and only
to defined the graph Laplacian, and all subsequent uses of ``adjacent'' and ``adjacency'' pertain
to differential privacy (not the adjacency matrix).}. 
\begin{definition}\label{dfn:adjacency}
    Let~$A \in \mathbb{N}$ be given, and
    fix a number of nodes~$n \in \mathbb{N}$. Two graphs on~$n$ nodes,~$G$ and~$G'$, are adjacent
    if they differ by $A$ edges. We express this mathematically via
    \begin{equation}
    \adj_A(G, G') = \begin{cases} 1 & |E(G) \Delta E(G')| \leq A \\
                           0 & \textnormal{otherwise}
               \end{cases},
    \end{equation}
    where~$S_1 \Delta S_2 = (S_1 \backslash S_2) \cup (S_2 \backslash S_1)$ is the symmetric difference of two sets
    and~$|\cdot|$ denotes cardinality. \hfill $\lozenge$
\end{definition}

Thus,~$A$ is the number of edges whose absence or presence must be concealed by privacy. In other words,
differential privacy for~$\lambda_2$ must make any graph approximately indistinguishable from any graph within~$A$ edges of it. 

Next, we briefly review differential privacy; see~\cite{dwork_algorithmic_2013} for a complete exposition.
A privacy mechanism $\mathcal{M}$ for a function $f$ is obtained by 
first computing the function $f$ on a given input $x$, and then adding noise to the output. 
The distribution of noise depends on the sensitivity of the function $f$ to changes in its input, described below.
It is the role of a mechanism to approximate functions
of sensitive data with private responses, and we next state this formally. 

\begin{definition}[\emph{Differential privacy; \cite{dwork_algorithmic_2013}}]\label{dfn:dp}
    Let $\epsilon > 0$ and $\delta \in [0,1)$ be given and fix a probability space $(\Omega, \mathcal{F}, \mathbb{P})$. 
    Then a mechanism $\mathcal {M}: \Omega \times \mathcal{G}_n \rightarrow \mathbb{R}$ is $(\epsilon , \delta)$-differentially private if,
    for all adjacent graphs $G,G' \in \mathcal{G}_n$, 
    \begin{equation}
        \mathbb{P}\big[\mathcal{M}(G) \in S\big] \leq \exp (\epsilon) \cdot \mathbb{P}\big[\mathcal{M}\left(G'\right) \in S\big] + \delta
    \end{equation}
    for all sets $S$ in the Borel $\sigma$-algebra over $\mathbb{R}$. \hfill $\lozenge$
\end{definition}

The value of~$\epsilon$ controls the amount of information shared, and typical values range from~$0.1$ to~$\log 3$~\cite{dwork_algorithmic_2013}.
The value of~$\delta$ can be regarded as the probability that more information is shared than~$\epsilon$ should allow, and typical
values range from~$0$ to~$0.05$. Smaller values of both imply stronger privacy. 
Given~$\epsilon$ and~$\delta$, a privacy mechanism must enforce Definition~\ref{dfn:dp} for all
graphs adjacent in the sense of Definition~\ref{dfn:adjacency}. 

We next define the sensitivity of~$\lambda_2$, which will be used later to calibrate the variance of privacy noise.  
With a slight abuse of notation, we treat~$\lambda_2$ as a function~$\lambda_2: \mathcal{G}_n \to \mathbb{R}$. 
\begin{definition}\label{dfn:sensitivity}
    The sensitivity of~$\lambda_2$ is the greatest difference
    between its values on Laplacians of adjacent graphs. Formally, given~$A$,
    \begin{equation}
    \Delta \lambda_2 = \max_{\substack{G, G' \in \mathcal{G}_n \\ \adj_A(G, G') = 1}} \big|\lambda_2(L) - \lambda_2(L')\big|,
    \end{equation}
    where $L$ and $L'$ are the Laplacians of $G$ and $G'$. \hfill $\lozenge$
\end{definition}

We next state the problems that we will solve. 

\begin{problem} \label{prob:one}
Given the adjacency relation in Definition~\ref{dfn:adjacency}, develop a mechanism to provide $(\epsilon, \delta)$-differentially privacy for the algebraic connectivity of a graph $G$.
\end{problem}

\begin{problem} \label{prob:two}
Given a diferentially private algebraic connectivity, 
quantify the accuracy of consensus protocol convergence rate estimates that use it. 
\end{problem}

\begin{problem} \label{prob:three}
    Given a private algebraic connectivity, develop bounds on the expectation of other graph properties.
\end{problem}

\subsection{Example Graph Privacy Attacks}
We close this section with two 
specific privacy attacks to highlight 
the importance of privacy
for the algebraic connectivity of graphs.
In each one, a single node combines knowledge
of its neighborhood set with knowledge
of~$\lambda_2$ to make inferences about a
graph's topology. There exist many possibilities
for more sophisticated attacks by considering
collusion among nodes to make inferences, and these
examples illustrate only two basic possibilities. 
%
%

\begin{example}
Consider Figure~\ref{fig:example3graph}, where node~$1$ wishes to infer node~$4$'s neighbors
using 
its own neighborhood set, i.e.,~$\mathcal{N}_1 = \{2, 3\}$, and
the fact that there are~$n=4$ nodes.

\begin{figure}[H]
\centering
\includegraphics[scale=0.5]{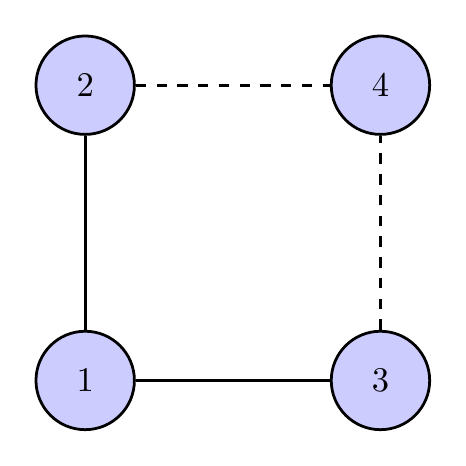}
\caption{Node~$1$ knows there are~$n=4$ nodes in the graph, and
it can use the value of~$\lambda_2$ to infer the connections of node~$4$. 
The dashed edges indicate edges that are present but unknown to agent~$1$. 
}
\label{fig:example3graph}
\end{figure}

The release of the value of~$\lambda_2$ provides node~$1$
with knowledge that~$\lambda_2 = 2$. 
Let~$d_{min}$ denote the minimum degree of the graph. Then, with the inequality~\cite{abreu07}
\begin{equation}
\lambda_2 \leq \frac{nd_{min}}{n-1},
\end{equation}
where~$n$ is the number of nodes, node~$1$ can infer that
\begin{equation}
d_{min} \geq \frac{(n-1)\lambda_2}{n} = \frac{3}{2}.
\end{equation}
Because~$d_{min}$ is integer-valued, we see that~$d_{min} \geq 2$.
Then node~$4$ must have at least two neighbors; because node~$4$ does
not share an edge with node~$1$, node~$1$ can infer, with certainty, that~$\mathcal{N}_4 = \{2, 3\}$. 
\hfill $\blacktriangle$
\end{example}

\begin{example}
Consider Figure~\ref{fig:example4graph}, and suppose that node~$1$ wishes to determine as much as possible with its neighborhood set,~$\mathcal{N}_1 = \{2, 3\}$, and the knowledge that~$\lambda_2 = 1$. 
\begin{figure}[H]
\centering
\includegraphics[scale=0.5]{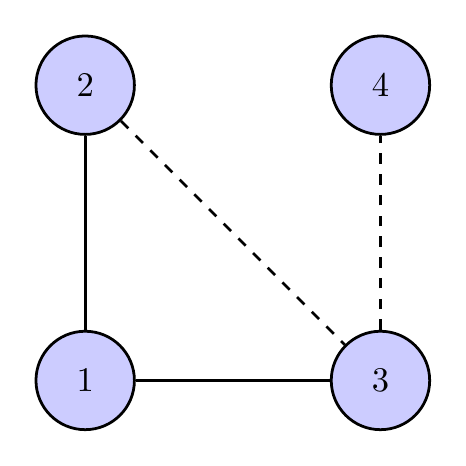}
\caption{Node~$1$ knows there are~$n=4$ nodes in the graph, and it can use the value of~$\lambda_2$ to essentially infer
the entire graph.
The dashed edges indicate edges that are present but unknown to agent~$1$. 
}
\label{fig:example4graph}
\end{figure}

From node~$1$'s perspective, 
the possible edges 
are~$E_{poss} = \{(2, 3), (2, 4), (3, 4)\}$, and node~$1$
wishes to determine which ones are present and absent.
Thus, there are~$8$ topologies to consider. 
Node~$1$ can rule out the case in which
all edges in~$E_{poss}$ are in the graph; if they
were, then the graph would have 
a ring graph as a sub-graph and
hence have~$\lambda_2 \geq 2$. 
If none
of the edges in~$E_{poss}$ were present, then the graph would be
disconnected, and it would have~$\lambda_2 = 0$.
Then either one or two edges from~$E_{poss}$ are present.

From~$E_{poss}$, if only~$(2, 3)$ were in the graph, then
it would also be disconnected, and thus node~$1$ can rule that
case out. If either~$(2, 4)$ or~$(3, 4)$ were in the graph
(with the others in~$E_{poss}$ absent), then it would have a line
topology, but this would give~$\lambda_2 = 2 - \sqrt{2}$.

Then exactly two edges from~$E_{poss}$ must be in the graph.
If both~$(2, 4)$ and~$(3, 4)$ were present
and~$(2, 3)$ were absent, then the graph would have
a ring topology, but then it would have~$\lambda_2 = 2$. 
Thus, the possibilities are either
that the edges~$(2, 3)$ and~$(3, 4)$ are in the graph,
or that the edges~$(2, 3)$ and~$(2, 4)$ are in the graph.

Then node~$1$ can conclude with certainty that
the edge~$(2, 3)$ is in the graph. It can also conclude
with certainty that either~$(2, 4)$ is in the graph
or~$(3, 4)$ is in the graph, but not both; the graphs
produced by having one of these edges
are isomorphic and hence cannot be distinguished here.
Thus, node~$1$ has inferred the topology and narrowed
the graph down to two possibilities for node labels
in that topology. 
\hfill $\blacktriangle$
\end{example}

We stress that these examples are only a small, representative sample of the kinds of privacy
attacks one can enact with knowledge of~$\lambda_2$. 
Broadly speaking, these are \emph{reconstruction attacks}, in that they combine released
information, namely~$\lambda_2$, with other knowledge to infer sensitive information,
which in this case is the underlying graph and/or its characteristics. There are many possibilities for
other knowledge of graph properties that can be combined with~$\lambda_2$, and, given the extensive suite of relationships
between~$\lambda_2$ and other graph properties~\cite{abreu07}, many attacks
are possible. 

Related graph privacy threats have been observed in the data science community
for other scalar-valued graph properties, such as counts of subgraphs and triangles~\cite{Ding2018}, degree sequences~\cite{Hay2009},
and numerous others~\cite{Task2012}. These threats have been addressed by developing new mechanisms to provide
differential privacy to the graph properties of interest. 
Given the privacy threats associated with releasing~$\lambda_2$ 
and the wide use of~$\lambda_2$ in analyzing multi-agent systems, we develop techniques to protect~$\lambda_2$ with differential privacy.

\section{Privacy mechanism for $\lambda_2$}\label{sec:results}
In this section we develop the privacy mechanism that enforces 
edge differential privacy.  We first start by providing a bound on the sensitivity in Definition~\ref{dfn:sensitivity}.
\begin{lemma}\label{lem:sensitivity_ub}
Fix an adjacency parameter~$A \in \mathbb{N}$. Then, with respect
to the adjacency relation in Definition~\ref{dfn:adjacency}, 
the sensitivity of $\lambda_2$ satisfies 
$    \Delta \lambda_2 \le 2A$.
\end{lemma}
\emph{Proof:} See Appendix~\ref{sec:adjacencyproof}.\hfill $\blacksquare$

Noise is added by a mechanism, which is a randomized map used to implement differential privacy. 
The Laplace mechanism is widely used, and it adds noise from a Laplace distribution to sensitive data (or functions thereof). 
The standard Laplace mechanism has support on all of~$\mathbb{R}$, though for graphs on~$n$ nodes,~$\lambda_2$ is known to lie in the interval~$[0, n]$. 
One can add Laplace noise and then project the result onto~$[0, n]$ (which is differentially private because the projection is merely post-processing), though
similar approaches have been shown to produce highly inaccurate private data~\cite{gohari21}. 
Instead, we use the bounded Laplace mechanism in~\cite{holohan2018bounded}; though bounded Laplace
noise appeared earlier in the privacy literature, to the best of our knowledge~\cite{holohan2018bounded} is the
first work to rigorously analyze its privacy properties. 
We state it in a form amenable to use with~$\lambda_2$. 

\begin{definition}    \label{dfn:bounded_laplace}
 Let $b>0$ and let $D = [0, n]$. 
	Then the bounded Laplace mechanism $W_{\lambda_2}:\Omega \rightarrow D$, for each
	$\lambda_2\in D$, is given by its probability density function $f_{W_{\lambda_2}}$ as
	\begin{equation}
		f_{W_{\lambda_2}}(x)
		=\begin{cases}
			0 & \text{if }x\notin D\\
			\frac{1}{C_{\lambda_2}(b)}\frac{1}{2b}e^{-\frac{|x-\lambda_2|}{b}} & \text{if }x\in D
		\end{cases},
	\end{equation}
	\noindent where $C_{\lambda_2}(b)=\int_{D}\frac{1}{2b}e^{-\frac{|x-\lambda_2|}{b}}dx$ is a normalizing
term. \hfill $\lozenge$
\end{definition}

Next, we establish an algebraic relation for $b$ which lets the bounded Laplace mechanism satisfy the theoretical guarantees  of $(\epsilon, \delta)$-differential privacy in Definition~\ref{dfn:dp}.
\begin{theorem}
[\emph{Privacy mechanism for~$\lambda_2$; Solution Problem~\ref{prob:one}}]
\label{thm:mechanim_variance_ub}
Let~$\epsilon > 0$ and~$\delta \in (0, 1)$ be given. Then for the bounded Laplace mechanism~$W_{\lambda_2}$ in 
Definition~\ref{dfn:bounded_laplace}, choosing $b$ according to 
\begin{equation} \label{eq:bdef}
    	b\ge\frac{2A}{\epsilon-\log\left(\frac{1-\frac{1}{2}e^{-\frac{2A}{b}}\left(1+e^{-\frac{n}{b}-1}\right)}{1-\frac{1}{2}\left(1+e^{-\frac{n}{b}}\right)}\right)-\log(1-\delta)}
\end{equation}
satisfies $(\epsilon , \delta)$-differentially privacy.
\end{theorem}
\emph{Proof:} See Appendix~\ref{sec:mechanismproof}.\hfill $\blacksquare$

Note that $b$ appears on both sides of~\eqref{eq:bdef}. 
In \cite{holohan2018bounded}, the authors provide an algorithm to solve for~$b$ using the bisection method, and we use this in the remainder of the paper.

\section{Applications to Consensus}
In this section, we solve Problem~\ref{prob:two} and bound the error in consensus convergence rates when they are computed using private values of~$\lambda_2$. 
As discussed in the introduction, consensus protocols underlie a number of multi-agent control and optimization algorithms, e.g.,~\cite{ren07,olfati04,gennaro06,nedic18}. 
Consider a network of~$n$ agents running a consensus protocol with communication topology modeled by an undirected, unweighted graph $G$. 
To protect the connections in this graph, a
differentially private version of~$\lambda_2$ is used for analysis. 

In continuous time, a consensus protocol takes the form $\dot{x}=-L(G)x,$ where $L(G)$ is the graph Laplacian. This protocol converges 
to the average of agents' initial state values with error bound at time~$t$ proportional to~$e^{-\lambda_2 t}$ \cite{Mesbahi2010}. 
Let $r(t)=e^{-\lambda_2 t}$ be the true convergence rate for the network $G$. 
Let $\tilde{\lambda}_2$ be the output of the bounded Laplace mechanism with privacy parameters $\epsilon$ and $\delta$. 
Let $\tilde{r}(t)=e^{-\tilde{\lambda}_2 t}$ be the convergence rate estimate. To compare the estimated convergence rate 
under privacy to the true convergence rate, we analyze $|\tilde{r}(t)-r(t)|$. 

Note that as $t\to \infty$, both~$\tilde{r}(t)\to 0$ and $r(t)\to 0$, which implies that $|\tilde{r}(t)-r(t)|\to 0$ as well. 
Although the error in the convergence rate estimate goes to~$0$ asymptotically, we are interested in analyzing the error 
at all values of $t$. 
To accomplish this, we give a concentration bound that bounds the 
probability $P\left(|\tilde{r}(t)-r(t)|\geq a\right)$ in terms of~$t$, the true algebraic connectivity~$\lambda_2$, and the level of the privacy encoded in $b$ that is determined using $\epsilon$ and $\delta$.

\begin{theorem}[\emph{Convergence rate concentration bound; Solution to Problem 2}] \label{thm:consensus}
Let $C_{\lambda_2}(b)=1-\frac{1}{2}\left(e^{-\frac{\lambda_{2}}{b}}+e^{-\frac{n-\lambda_{2}}{b}}\right)$. 
Then, for $t > 0$ and a fixed $\lambda_{2}$ and $b$,
\begin{equation}
P\left(|\tilde{r}(t)-r(t)|\geq a\right)\leq \frac{1}{C_{\lambda_2}(b)}\frac{1}{2a}\left(\rho_1(t)+\rho_2(t)-\rho_3(t)\right),
\end{equation}
where
\begin{align}
    \rho_1(t) &= \frac{e^{-\lambda_{2}(\frac{1}{b}+t)}\left(-bte^{\frac{\lambda_{2}}{b}}+bt+e^{\lambda_{2}t}-1\right)}{bt-1}, \\
    \rho_2(t) \!&=\! e^{-\lambda_{2}t}\left(1-e^{\frac{\lambda_{2}-n}{b}}\right)\!, \,\,
    \rho_3(t) \!=\! \frac{e^{-\lambda_{2}t}-e^{\frac{\lambda_{2}-n(bt+1)}{b}}}{bt+1}.
\end{align}
\end{theorem}

\emph{Proof:} See Appendix \ref{sec:thm2proof}.
 \hfill $\blacksquare$

Taking limits of the bound presented in Theorem~\ref{thm:consensus} shows that as $t\to\infty,$ $P\left(|\tilde{r}(t)-r(t)|\geq a\right)\to 0$ for all $a$, and thus this bound has the expected asymptotic behavior. 

We can use Theorem~\ref{thm:consensus} to further characterize the transient response of error in the estimated
consensus convergence rate. 
Specifically, we can bound the 
time required for the error in the convergence rate estimate to be 
larger than some threshold~$a$ only with probability smaller than $\eta$. Formally, 
given a threshold~$a > 0$ and probability~$\eta > 0$,
we bound the times~$t$ for which~$P\left(|\tilde{r}(t)-r(t)|\geq a\right)\leq  \eta$. 

\begin{theorem} \label{thm:times}
Fix~$a > 0$ and~$\eta \in (0, 1)$. 
Let~$\epsilon > 0$ and~$\delta \in (0, 1)$ be given and compute the scale parameter~$b > 0$. 
Consider a graph on~$n$ nodes with algebraic connectivity~$\lambda_2$.
If~$\lambda_2 \leq \frac{n}{2}$, then 
we have~$P\left(|\tilde{r}(t)-r(t)|\geq a\right)\leq\eta$
for 
\begin{equation}
t\geq\frac{\left(e^{-\frac{\lambda_{2}}{b}}-e^{\frac{\lambda_{2}-n}{b}}\right)\frac{b}{\lambda_{2}e}+2aC_{\lambda_2}(b)\eta+1}{2aC_{\lambda_2}\eta b}. 
\end{equation}
If~$\lambda_{2}>\frac{n}{2}$, then the desired bound holds for~$t\geq\frac{2aC_{\lambda_2}(b)\eta+1}{2aC_{\lambda_2}(b)\eta b}$. 
\end{theorem}
\emph{Proof:} 
See Appendix~\ref{sec:timebound}.
\hfill $\blacksquare$

We note that the statistics of the differential privacy mechanism can be released
without harming privacy. Therefore, the values of~$C_{\lambda_2}(b)$ and~$b$
can be publicly released. A network analyst can compute these bounds
for any choices of~$a$ and~$\eta$ of interest. Because the exact
value of~$\lambda_2$ is unknown, they can compute the maximum
value of these two times to find a time after which the
desired error bound always holds.

Notice that the two conditions on $t$ in Theorem~\ref{thm:times} 
only vary by a factor of $\left(e^{-\frac{\lambda_{2}}{b}}-e^{\frac{\lambda_{2}-n}{b}}\right)\frac{b}{\lambda_{2}e}$ in the numerator, 
and
when $\lambda_2$ is large this term is negative. 
This means that if $\lambda_2$ is large, the required time for~$P\left(|\tilde{r}(t)-r(t)|\geq a\right)\leq\eta$ is smaller than if $\lambda_2$ was small. 
In Section~\ref{sec:casestudy}, we provide simulation results and further commentary for Theorem 1.

Beyond the consensus protocol, the value of~$\lambda_2$ is related to many other
graph properties~\cite{abreu07}, and we next show how the private value of~$\lambda_2$
can still be used to accurately bound two other properties of interest.

\section{Bounding Other Graph Properties}
There exist numerous inequalities relating~$\lambda_2$ to
other quantitative graph properties~\cite{abreu07,Mesbahi2010}, 
and one can therefore expect that the private~$\lambda_2$ will be used to estimate
other quantitative characteristics of graphs. 
To illustrate the utility of doing so, 
in this section we bound the
graph diameter~$d$ and mean distance~$\rho$
in terms of the private value~$\tilde{\lambda}_2$. 

Both~$d$ and~$\rho$ measure graph size and provide insight into 
how easily information can be transferred across a network~\cite{PALDINO2017201}. 
We estimate each one in terms of the private~$\lambda_2$ and bound the error
induced in these estimates by privacy. 
These bounds represent
Similar bounds can be simply derived, e.g., on minimal/maximal degree, 
edge connectivity, etc., because their bounds are proportional to~$\lambda_2$~\cite{fiedler_algebraic_1973}.

We first recall bounds from the literature. 
\begin{lemma}[Diameter and Mean Distance Bounds\cite{Mohar1991eigenvalues}]\label{thm:d_rho_bounds}
    For an undirected, unweighted graph $G$ of order $n$, define 
    \begin{align*}
        &\overline d(\lambda_2,\alpha) =  \left(2\sqrt{\frac{\lambda_n}{\lambda_2}}\sqrt{\frac{\alpha^2-1}{4\alpha}}+2\right)\left(\log_\alpha\frac{n}{2}\right) \\
        &\overline\rho(\lambda_2,\alpha) =  \left(\!\sqrt{\frac{\lambda_n}{\lambda_2}}\sqrt{\frac{\alpha^2-1}{4\alpha}}\!+\!1\!\right)\left(\!\frac{n}{n-1}\!\right)\left(\frac{1}{2} \!+\! \log_\alpha\frac{n}{2}\right).
    \end{align*}
    
    Then for any fixed~$\lambda_2>0$ and any~$\alpha>1$,
    the diameter~$d$ and mean distance~$\rho$ of the graph~$G$ are bounded via
    \begin{align}
        &\underline{d}(\lambda_2)=\frac{4}{n\lambda_2}\leq d \leq  \overline d(\lambda_2,\alpha) \\
        &\underline{\rho}(\lambda_2)=\frac{2}{(n-1)\lambda_2} + \frac{n-2}{2(n-1)} \leq \rho \leq \overline\rho(\lambda_2,\alpha). 
    \end{align}
    
    The least upper bounds can be derived by using $\alpha_d$ and $\alpha_\rho$ which minimize $\overline d(\lambda_2,\alpha)$ and $\overline\rho(\lambda_2,\alpha)$ respectively.\hfill $\blacksquare$
\end{lemma} 

A list of $\alpha_d$ and $\alpha_\rho$ can be found in Table 1 in \cite{Mohar1991eigenvalues}. To quantify the impacts of 
using the private~$\lambda_2$ in these bounds, we next bound the expectations of~$d$ and~$\rho$. 
These bounds use the upper incomplete gamma function~$\Gamma(\cdot,\cdot)$ and the imaginary error function~$\textrm{erfi}(\cdot)$,
defined as
\begin{equation}
\Gamma(s,x) = \int_x^\infty t^{s-1}e^{-t}dt \,\,\,\textnormal{ and }\,\,\, \textrm{erfi}(x) = \frac{2}{\sqrt{\pi}}\int_0^x e^{t^2}dt.
\end{equation}

Using the private~$\lambda_2$, expectation bounds are as follows. 
\begin{theorem}[\emph{Expectation bounds for~$d$ and~$\rho$; Solution to Problem 3}]\label{thm:expectation_bounds_d_rho}
    For any $\lambda_2> 0$, denote its private value by~$\tilde{\lambda}_2$. 
    Then, when bounded using~$\tilde{\lambda}_2$, the expectations of the diameter,~$E[\tilde{d}]$, and mean distance,~$E[\tilde{\rho}]$, obey
    \begin{align}
        &\frac{4}{nE[\tilde{\lambda}_2]}\leq E[\tilde{d}] \leq E[\overline{d}(\tilde{\lambda}_2,\alpha_d)] \qquad \textnormal{ and }\\
        &\frac{2}{(n-1)E[\tilde{\lambda}_2]} + \frac{n-2}{2(n-1)} \leq E[\tilde{\rho}]\leq E[\overline{\rho}(\tilde{\lambda}_2,\alpha_\rho)],
    \end{align}
    where
    \begin{align*}
        &E[\overline{d}(\tilde{\lambda}_2,\alpha_d)] \!=\!\left[2\sqrt{\frac{\lambda_n(\alpha_d^2-1)}{4\alpha_d}}E\!\left[\sqrt{\frac{1}{\tilde{\lambda}_2}}\right] \!+\! 2\right]\!\!\left[\log_{\alpha_d}\frac{n}{2}\right] \\
        &\begin{multlined}E[\overline{\rho}(\tilde{\lambda}_2,\alpha_\rho)]=\left[\sqrt{\frac{\lambda_n(\alpha_\rho^2-1)}{4\alpha_\rho}}E\left[\frac{1}{\sqrt{\tilde{\lambda}_2}}\right]+1\right]\\
        \cdot\left[\frac{n}{n-1}\right]\cdot\left[\frac{1}{2}+\log_{\alpha_\rho}\frac{n}{2}\right]\end{multlined}.
    \end{align*}
    We can compute the expectation terms with~$\tilde{\lambda}_2$ via
    \begin{align*}
        &\begin{multlined}
        E\left[\frac{1}{\sqrt{\tilde{\lambda}_2}}\right] = \frac{1}{C_{\lambda_2}(b)}\frac{1}{2b}\left(\sqrt{\pi}\sqrt{b}e^{-\frac{\lambda_2}{b}}\left(\textrm{erfi}\left(\sqrt{\frac{\lambda_2}{b}}\right)\right)\right.\\
        \left.+\sqrt{b}e^{\frac{\lambda_2}{b}}\left(\Gamma\left(\frac{1}{2},\frac{n}{b}\right)-\Gamma\left(\frac{1}{2},\frac{\lambda_2}{b}\right)\right)\right)
        \end{multlined} \\
        &\begin{multlined}
        E[\tilde{\lambda}_2] = \frac{1}{2C_{\lambda_2}(b)}\left(2\lambda_2+be^{-\frac{\lambda_2}{b}}-be^{-\frac{n-\lambda_2}{b}}-ne^{-\frac{n-\lambda_2}{b}}\right)
        \end{multlined}
    \end{align*}
\end{theorem}

\emph{Proof:} 
See Appendix~\ref{apdx:thm4}.
\hfill $\blacksquare$

\begin{remark}\label{rmk:epsilon_bounds_relation}
    A larger~$\epsilon$ indicates weaker 
    privacy, and it results in a smaller value of $b$ and a distribution
    of privacy noise that is more tightly concentrated about its mean. 
    Thus, a larger~$\epsilon$ implies that the expected value
    $E[\tilde{\lambda}_2]$ is closer to the exact,
    non-private $\lambda_2$, which leads to smaller disagreements in the bounds on the exact and expected values of $d$ and $\rho$.     
\end{remark}


\section{Simulations}\label{sec:casestudy}
 In this section, we present consensus simulation results and 
 numerical results for
 the bounds on graph measurements when using the private~$\lambda_2$ in graph analysis. 

Consider a network of $n=10$ agents with $\lambda_2=1$ and a true convergence rate of $r(t)=e^{-\lambda_2 t} = e^{-t}$.  
The network operator wishes to privatize~$\lambda_2$ 
 using the bounded Laplace mechanism with $\epsilon=0.4,\ \delta=0.05,$ and $A=1.$ 
 Solving for $b$ with the algorithm in \cite{holohan2018bounded} yields $b\geq7.39,$ and selecting $b=7.39$ ensures $(0.4,0.05)-$differential privacy. Let $\tilde{\lambda}_2$ 
 be the private output of the bounded Laplace mechanism. 
 Then, for a recipient of~$\tilde{\lambda}_2$, 
 the estimated consensus convergence rate is~$\tilde{r}(t)=e^{-\tilde{\lambda}_2 t}$. 
 Let $P\big(|\tilde{r}(t)-r(t)|\leq a\big)$ be the the probability of the error of the convergence rate estimate being less than $a$ at time $t$. 
 Intuitively, $P\big(|\tilde{r}(t)-r(t)|\leq a\big)$ should be close to $1$ as~$t$ grows. 
 
\begin{figure}
    \centering
    \includegraphics[width=.45\textwidth]{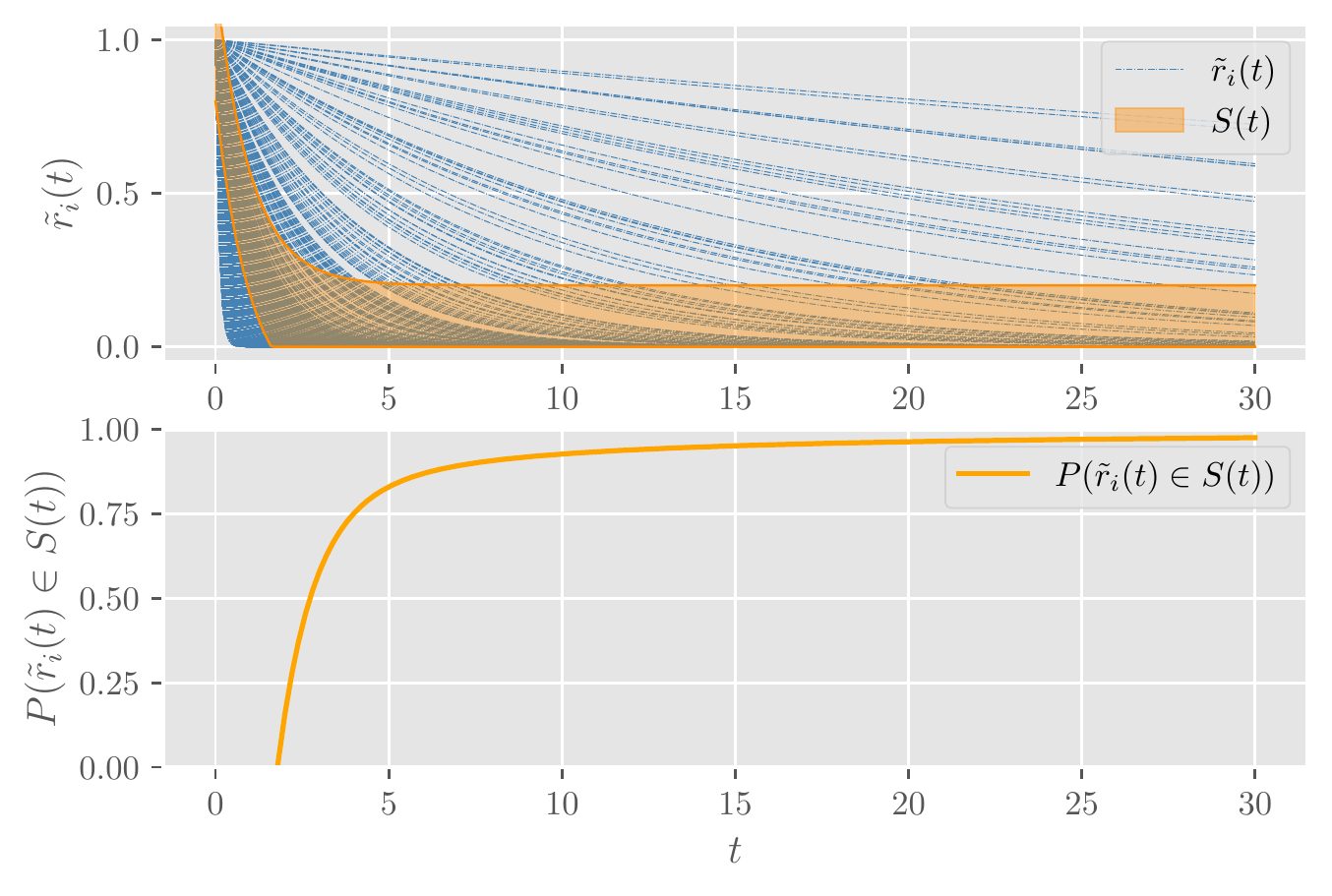}
    \vspace{-5mm}
    \caption{The top plot shows $500$ sample convergence rate estimates~$\tilde{r}_i(t)=e^{-\tilde{\lambda}_2^i t}$, where~$\tilde{\lambda}_2^i$ is the output of 
    the~$i^{th}$ trial of the bounded Laplace mechanism with $\epsilon=0.4,\ \delta=0.05,$ and $A=1$ for a network of $n=10$ agents with $\lambda_2=1.$ The set $S(t)$ 
    shown on these plots is defined as~${S(t) = \{\tilde{r}_i(t) : \tilde{r}_i(t) = e^{-\tilde{\lambda}_2^it} \textnormal{ and } |\tilde{r}_i(t)-e^{-\lambda_2t}|\leq a\}}$ with $a=0.2.$ The bottom plot shows the lower bound on $P(\tilde{r}_i(t)\in S)$ obtained by using Theorem~1 in~\eqref{eq:simlowerbound}. 
    This lower bound approaches $1$ relatively quickly and is consistent with the sample convergence rates shown in the top plot.}
    \label{fig:cal_sim}
\end{figure}

 We can lower bound~$P\big(|\tilde{r}(t)-r(t)|\leq a\big)$ by noting that 
 \begin{equation} \label{eq:simlowerbound}
 P\big(|\tilde{r}(t)-r(t)|\leq a\big)=1-P\big(|\tilde{r}(t)-r(t)|\geq a\big), 
 \end{equation}
 which we can use Theorem~\ref{thm:consensus} to bound. 
 To that end, 
 Figure~\ref{fig:cal_sim} shows how~$P\big(|\tilde{r}(t)-r(t)|\leq a\big)$ 
 changes with time for $a=0.2$ and shows $500$ sample convergence rate estimates 
 for values of~$\lambda_2$ privatized with the parameters from above.

These simulations show that for a sufficiently large~$t$, $P\big(|\tilde{r}(t)-r(t)|\leq a\big)$ is close to $1$ and that the times $t$ 
for which this occurs are often small. 
This occurs because the bounded Laplace mechanism outputs $\tilde{\lambda}_2 \in [0,n]$,
and thus~$\tilde{r}(t)\to 0$ as $t\to \infty$ for any~$\tilde{\lambda}_2$, while 
the true convergence rate~$r(t)$ also converges to $0$. These results also show that Theorem~1 is consistent with intuition as highlighted in 
Figure~\ref{fig:cal_sim}, namely that as $t$ grows the error in any estimated convergence rate using the output of the bounded Laplace mechanism converges to 0
eventually. 

\begin{figure}
    \centering
    \includegraphics[width=.45\textwidth]{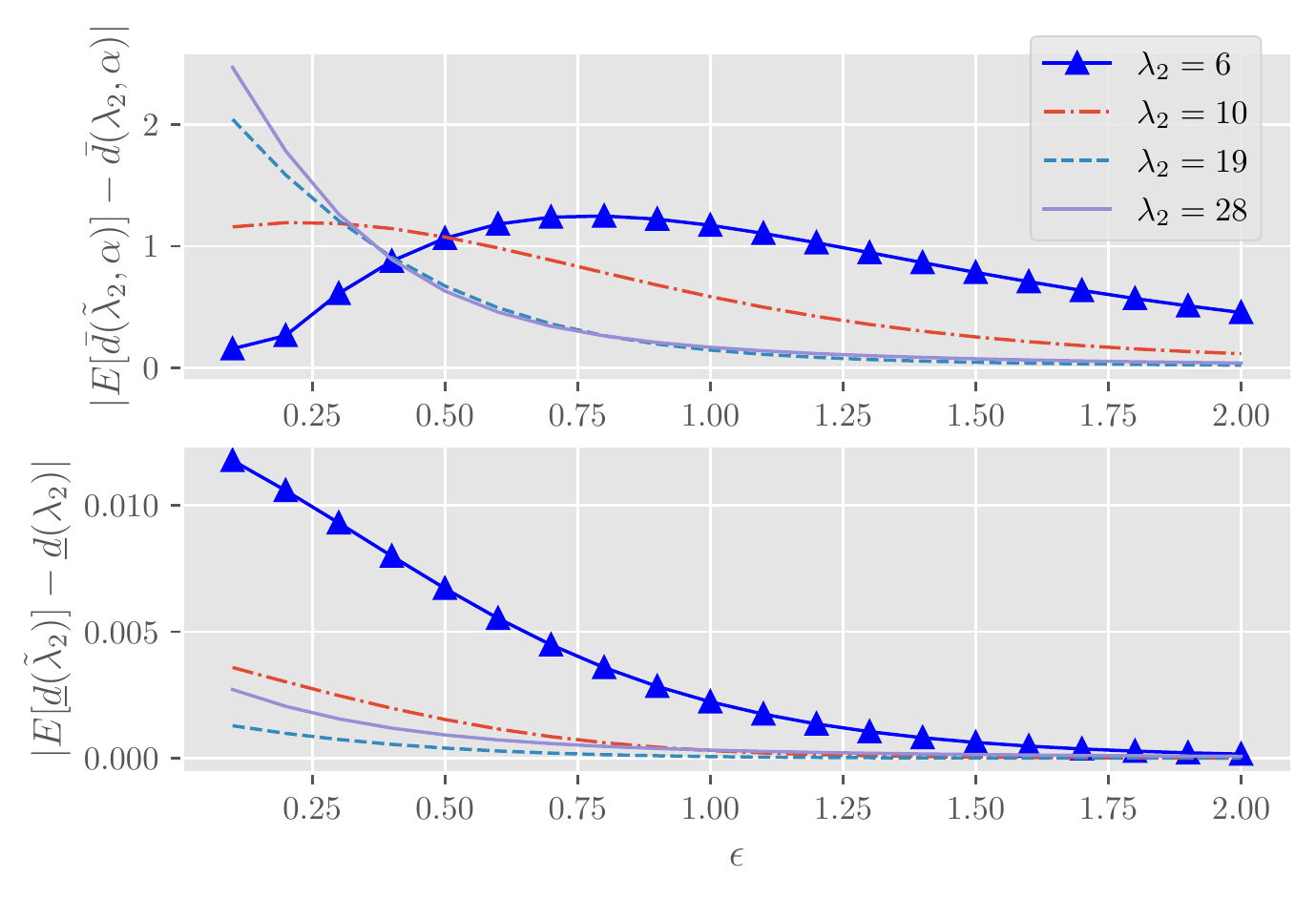}
    \vspace{-5mm}
    \caption{The top plot shows the distance between the exact and expected upper bounds for~$d$. The bottom one shows the distance between the corresponding lower bounds.}
    \label{fig:bounds_for_d}
\end{figure}

\begin{figure}
    \centering
    \includegraphics[width=.45\textwidth]{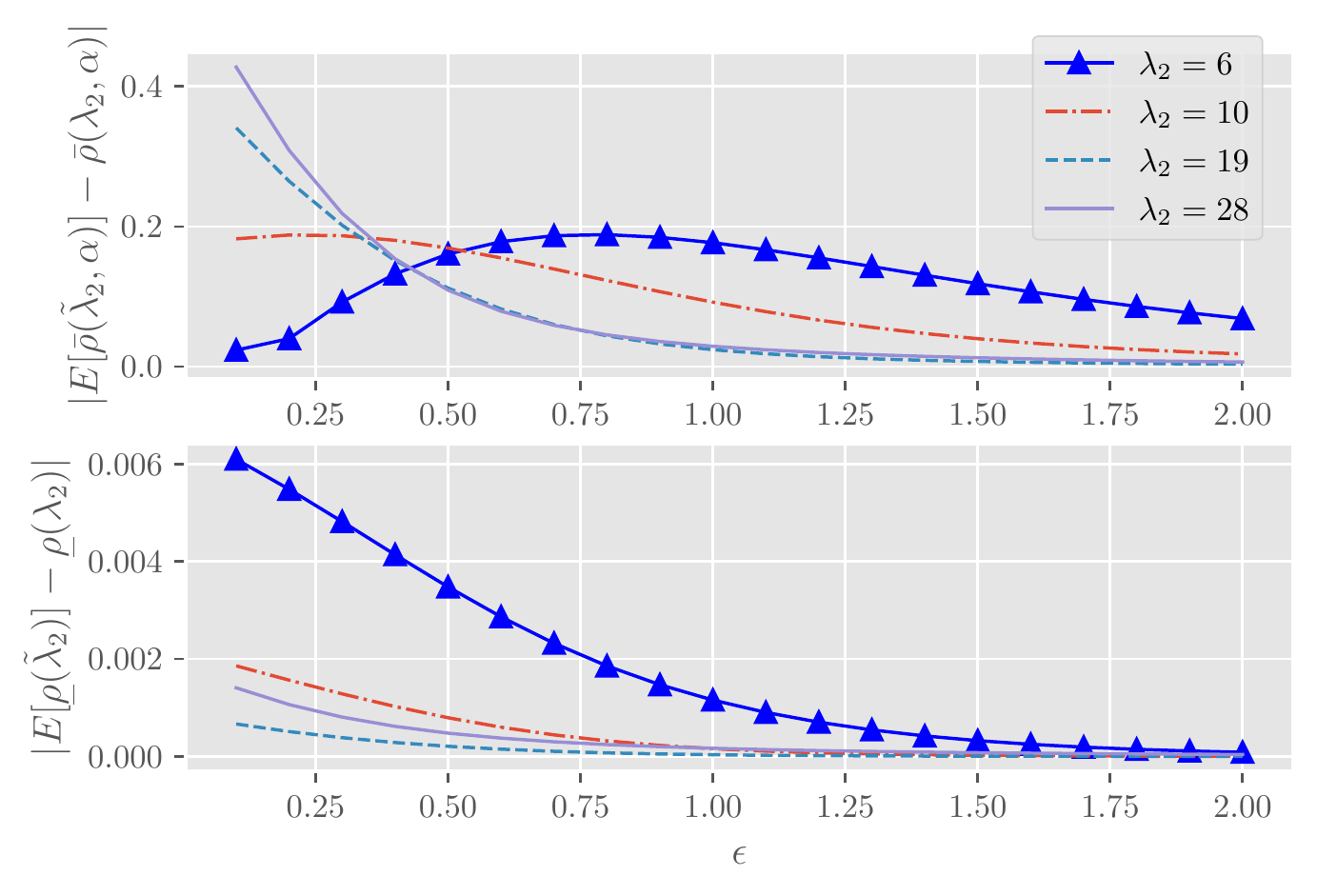}
    \vspace{-5mm}
    \caption{The top plot shows the distance between the exact and expected upper bounds for~$\rho$. The bottom one shows the distance between the corresponding lower bounds.}
    \label{fig:bounds_for_rho}
\end{figure}

We next present simulation results for using the private value of~$\lambda_2$ 
to estimate $d$ and $\rho$. We consider networks of $n=30$ agents with different edge sets 
and hence different values of~$\lambda_2$. We let $\lambda_n=n$  and therefore the 
upper bounds on $d$ and $\rho$ in Theorem~\ref{thm:expectation_bounds_d_rho} 
can reach their worst-case values. We apply the bounded 
Laplace mechanism with $\delta=0.05$ and a range of $\epsilon\in[0.1,2]$.
To illustrate the effects of privacy in bounding diameter, we compute the distance
between the exact (non-private) upper bound on diameter in Lemma~\ref{thm:d_rho_bounds}
and the expected (private) upper bound on diameter in Theorem~\ref{thm:expectation_bounds_d_rho}. 
This distance is shown in the upper plot in Figure~\ref{fig:bounds_for_d}, and the lower
plot shows the analogous distance for the diameter lower bounds. 
Figure~\ref{fig:bounds_for_rho} shows the corresponding upper- and lower-bound distances
for~$\rho$. 

In all plots, there is a general decrease in the distance between the exact and private bounds as~$\epsilon$ grows. 
Recalling that larger~$\epsilon$ implies weaker privacy, 
these simulations confirm that weaker privacy guarantees result in smaller differences between the exact and expected bounds for $d$ and $\rho$, as predicted in Remark~\ref{rmk:epsilon_bounds_relation}. 
\section{Conclusions} \label{sec:conclusions}
This paper presented a differential privacy mechanism for the
algebraic connectivity of undirected, unweighted graphs.
Bounded noise was used to provide private values that are still
accurate, and the private values of algebraic connectivity
were shown to give accurate estimates of consensus protocol
convergence rates, and the diameter and mean distance
of a graph. Future work includes the development of
new privacy mechanisms for other algebraic graph properties.


\bibliographystyle{IEEEtran}{}
\bibliography{sources}

\begin{thebibliography}{10}
\providecommand{\url}[1]{#1}
\csname url@samestyle\endcsname
\providecommand{\newblock}{\relax}
\providecommand{\bibinfo}[2]{#2}
\providecommand{\BIBentrySTDinterwordspacing}{\spaceskip=0pt\relax}
\providecommand{\BIBentryALTinterwordstretchfactor}{4}
\providecommand{\BIBentryALTinterwordspacing}{\spaceskip=\fontdimen2\font plus
\BIBentryALTinterwordstretchfactor\fontdimen3\font minus
  \fontdimen4\font\relax}
\providecommand{\BIBforeignlanguage}[2]{{%
\expandafter\ifx\csname l@#1\endcsname\relax
\typeout{** WARNING: IEEEtran.bst: No hyphenation pattern has been}%
\typeout{** loaded for the language `#1'. Using the pattern for}%
\typeout{** the default language instead.}%
\else
\language=\csname l@#1\endcsname
\fi
#2}}
\providecommand{\BIBdecl}{\relax}
\BIBdecl

\bibitem{Ren2005}
{Wei Ren}, R.~W. {Beard}, and E.~M. {Atkins}, ``A survey of consensus problems
  in multi-agent coordination,'' in \emph{Proceedings of the 2005, American
  Control Conference, 2005.}, 2005.

\bibitem{Scott1988}
J.~Scott, ``Social network analysis,'' \emph{Sociology}, vol.~22, no.~1, pp.
  109--127, 1988.

\bibitem{SHIRLEY2005287}
M.~D. Shirley and S.~P. Rushton, ``The impacts of network topology on disease
  spread,'' \emph{Eco. Complexity}, vol.~2, no.~3, pp. 287--299, 2005.

\bibitem{zheng_consensus_2011}
Y.~Zheng, L.~Wang, and Y.~Zhu, ``Consensus of heterogeneous multi-agent
  systems,'' vol.~5, no.~16, pp. 1881--1888.

\bibitem{Mieghem2009}
P.~{Van Mieghem}, J.~{Omic}, and R.~{Kooij}, ``Virus spread in networks,''
  \emph{IEEE/ACM Transactions on Networking}, vol.~17, no.~1, pp. 1--14, 2009.

\bibitem{freitas2020evaluating}
S.~Freitas and D.~H. Chau, ``Evaluating graph vulnerability and robustness
  using tiger,'' 2020.

\bibitem{kasiviswanathan13}
S.~P. Kasiviswanathan, K.~Nissim, S.~Raskhodnikova, and A.~Smith, ``Analyzing
  graphs with node differential privacy,'' in \emph{Proceedings of the 10th
  Theory of Cryptography Conference on Theory of Cryptography}.\hskip 1em plus
  0.5em minus 0.4em\relax Springer-Verlag, 2013, p. 457–476.

\bibitem{karwa14}
V.~Karwa, S.~Raskhodnikova, A.~Smith, and G.~Yaroslavtsev, ``Private analysis
  of graph structure,'' \emph{ACM Trans. Database Syst.}, vol.~39, no.~3, 2014.

\bibitem{kasiviswanathan14}
S.~P. Kasiviswanathan and A.~Smith, ``On the ’semantics’ of differential
  privacy: A bayesian formulation,'' \emph{Journal of Privacy and
  Confidentiality}, vol.~6, no.~1, Jun. 2014.

\bibitem{dwork_algorithmic_2013}
C.~Dwork and A.~Roth, ``The algorithmic foundations of differential privacy,''
  vol.~9, no.~3, pp. 211--407.

\bibitem{day16}
W.-Y. Day, N.~Li, and M.~Lyu, ``Publishing graph degree distribution with node
  differential privacy,'' in \emph{Proceedings of the 2016 International
  Conference on Management of Data}, 2016, p. 123–138.

\bibitem{shen13}
E.~Shen and T.~Yu, ``Mining frequent graph patterns with differential
  privacy,'' in \emph{Proceedings of the 19th ACM International Conference on
  Knowledge Discovery and Data Mining}, 2013, pp. 545--553.

\bibitem{olfati04}
R.~{Olfati-Saber} and R.~M. {Murray}, ``Consensus problems in networks of
  agents with switching topology and time-delays,'' \emph{IEEE Transactions on
  Automatic Control}, vol.~49, no.~9, pp. 1520--1533, 2004.

\bibitem{ren07}
W.~Ren and E.~Atkins, ``Distributed multi-vehicle coordinated control via local
  information exchange,'' \emph{International Journal of Robust and Nonlinear
  Control}, vol.~17, pp. 1002--1033, 2007.

\bibitem{gennaro06}
M.~C. De~Gennaro and A.~Jadbabaie, ``Decentralized control of connectivity for
  multi-agent systems,'' in \emph{Proceedings of the 45th IEEE Conference on
  Decision and Control}, 2006, pp. 3628--3633.

\bibitem{nedic18}
A.~Nedi{\'{c}}, A.~Olshevsky, and W.~Shi, \emph{Decentralized Consensus
  Optimization and Resource Allocation}, 2018, pp. 247--287.

\bibitem{holohan2018bounded}
N.~Holohan, S.~Antonatos, S.~Braghin, and P.~Mac~Aonghusa, ``The bounded
  laplace mechanism in differential privacy,'' \emph{arXiv preprint
  arXiv:1808.10410}, 2018.

\bibitem{yazdani2018differentially}
K.~Yazdani, A.~Jones, K.~Leahy, and M.~Hale, ``Differentially private lq
  control,'' \emph{arXiv preprint arXiv:1807.05082}, 2018.

\bibitem{yazdani2020error}
K.~Yazdani and M.~Hale, ``Error bounds and guidelines for privacy calibration
  in differentially private kalman filtering,'' in \emph{2020 American Control
  Conference (ACC)}, 2020, pp. 4423--4428.

\bibitem{hawkins2020differentially}
C.~Hawkins and M.~Hale, ``Differentially private formation control,'' in
  \emph{2020 59th IEEE Conference on Decision and Control (CDC)}, 2020.

\bibitem{gohari2020privacy}
P.~Gohari, M.~Hale, and U.~Topcu, ``Privacy-preserving policy synthesis in
  markov decision processes,'' in \emph{2020 59th IEEE Conference on Decision
  and Control (CDC)}, 2020.

\bibitem{wang13}
Y.~Wang, X.~Wu, and L.~Wu, ``Differential privacy preserving spectral graph
  analysis,'' in \emph{Pacific-Asia Conference on Knowledge Discovery and Data
  Mining}, 2013, pp. 329--340.

\bibitem{fiedler1975property}
M.~Fiedler, ``A property of eigenvectors of nonnegative symmetric matrices and
  its application to graph theory,'' \emph{Czechoslovak Mathematical Journal},
  vol.~25, no.~4, pp. 619--633, 1975.

\bibitem{abreu07}
N.~M.~M. {de Abreu}, ``Old and new results on algebraic connectivity of
  graphs,'' \emph{Linear Algebra and its Applications}, vol. 423, no.~1, pp.
  53--73, 2007.

\bibitem{Mesbahi2010}
M.~Mesbahi and M.~Egerstedt, \emph{Graph Theoretic Methods in Multiagent
  Networks}, 2010.

\bibitem{Ding2018}
X.~Ding, X.~Zhang, Z.~Bao, and H.~Jin, ``Privacy-preserving triangle counting
  in large graphs,'' in \emph{Proceedings of the 27th ACM International
  Conference on Information and Knowledge Management}.\hskip 1em plus 0.5em
  minus 0.4em\relax Association for Computing Machinery, 2018, p. 1283–1292.

\bibitem{Hay2009}
M.~{Hay}, C.~{Li}, G.~{Miklau}, and D.~{Jensen}, ``Accurate estimation of the
  degree distribution of private networks,'' in \emph{2009 Ninth IEEE
  International Conference on Data Mining}, 2009, pp. 169--178.

\bibitem{Task2012}
C.~{Task} and C.~{Clifton}, ``A guide to differential privacy theory in social
  network analysis,'' in \emph{International Conference on Advances in Social
  Networks Analysis and Mining}, 2012, pp. 411--417.

\bibitem{gohari21}
P.~{Gohari}, B.~{Wu}, C.~{Hawkins}, M.~{Hale}, and U.~{Topcu}, ``Differential
  privacy on the unit simplex via the dirichlet mechanism,'' \emph{IEEE
  Transactions on Information Forensics and Security}, vol.~16, pp. 2326--2340,
  2021.

\bibitem{PALDINO2017201}
M.~J. Paldino, W.~Zhang, Z.~D. Chu, and F.~Golriz, ``Metrics of brain network
  architecture capture the impact of disease in children with epilepsy,''
  \emph{NeuroImage: Clinical}, vol.~13, pp. 201--208, 2017.

\bibitem{fiedler_algebraic_1973}
M.~Fiedler, ``Algebraic connectivity of graphs,'' vol.~23.

\bibitem{Mohar1991eigenvalues}
B.~Mohar, ``Eigenvalues, diameter, and mean distance in graphs,'' \emph{Graph.
  Comb.}, 1991.

\bibitem{bernstein2009matrix}
D.~S. Bernstein, \emph{Matrix mathematics: theory, facts, and formulas}.\hskip
  1em plus 0.5em minus 0.4em\relax Princeton university press, 2009.

\end{thebibliography}
\appendix

\subsection{Proof of Lemma~\ref{lem:sensitivity_ub}}
\label{sec:adjacencyproof}
Consider two graphs $G, G' \in \mathcal{G}_n$ such that $\adj_A(G, G')=1$. Denote their corresponding graph Laplacians by $L$ and $L'$, and
define the matrix $P$ such that $L' = L+P$. Then, we write 
\begin{align}
 \Delta\lambda_{2} &= \max_{G,G' \in \mathcal{G}_n} \left|\lambda_{2}\left(L'\right)-\lambda_{2}\left(L\right)\right|
 \\
 &= \max_{G,G' \in \mathcal{G}_n} \left|\lambda_{2}\left(L+P\right)-\lambda_{2}\left(L\right)\right|. 
\end{align}
Applying \cite[Theorem 8.4.11]{bernstein2009matrix} to split up~$\lambda_2(L+P$), we obtain
\begin{align}
 \Delta\lambda_{2}& \le\lambda_{2}(L)+\lambda_{n}(P) - \lambda_{2}(L) = \lambda_n(P). 
\end{align}

The matrix~$P$ encodes the differences between~$L$ and~$L'$ as follows. For any~$i$, if the diagonal
entry~$P_{ii} = 1$, then node~$i$ has one more edge in~$G'$ than it does in~$G$. If~$P_{ii} = -1$, then
node~$i$ has one fewer edge in~$G'$ than it does in~$G$.
Other values of~$P_{ii}$ indicate the addition or removal of more edges. 
Given~$A$, we have~$-A \leq P_{ii} \leq A$. 

For off-diagonal entries,~$P_{ij} = 1$ 
indicates that~$G'$ contains the edge~$(i, j)$ and~$G$ does not; the converse holds
if~${P_{ij} = -1}$. Then, for any row of~$P$, the diagonal entry has absolute value at most~$A$,
and the absolute sum of the off-diagonal entries is at most~$A$. 
By Ger\v{s}gorin's circle theorem \cite[Fact 4.10.16.]{bernstein2009matrix}, we have $\lambda_n(P)\le 2A$. \hfill $\blacksquare$

\subsection{Proof of Theorem~\ref{thm:mechanim_variance_ub}}
\label{sec:mechanismproof}
By~\cite[Theorem 3.5]{holohan2018bounded}, the bounded Laplace
mechanism satisfies differential privacy if 
\begin{equation}
	b\ge\frac{\Delta \lambda_2}{\epsilon-\log\Delta C(b)-\log(1-\delta)},
\end{equation}
where, given that $\lambda_2\in [0,n]$, $ \Delta C(b)$ is defined as
\begin{equation}\label{eq:delta_C_b}
\Delta C(b) :=\frac{C_{\Delta \lambda_2}(b)}{C_{0}(b)}.
\end{equation}
Next, we compute the normalizing constant in Definition~\ref{dfn:bounded_laplace} as
\begin{equation} \label{eq:cl2}
	C_{\lambda_{2}}(b)=1-\frac{1}{2}\left(e^{-\frac{\lambda_{2}}{b}}+e^{-\frac{n-\lambda_{2}}{b}}\right).
\end{equation}
Using~\eqref{eq:cl2} in~\eqref{eq:delta_C_b} gives 
\begin{equation}
	\Delta C(b) =\frac{1-\frac{1}{2}\left(e^{-\frac{\Delta \lambda_2}{b}}+e^{-\frac{n-\Delta \lambda_2}{b}}\right)}{1-\frac{1}{2}\left(1+e^{-\frac{n}{b}}\right)}. 
\end{equation}
Using the sensitivity bound in Lemma~\ref{lem:sensitivity_ub}, we put $\Delta \lambda_2=2A$, which completes the proof. \hfill $\blacksquare$

\subsection{Proof of Theorem~\ref{thm:consensus}}
\label{sec:thm2proof}
Let $\tilde{\lambda}_{2}$ be the output of the bounded Laplace mechanism with scale parameter~$b$. We begin by computing
the expected value of $g(\tilde{\lambda}_{2})=|\tilde{r}(t)-r(t)|$
with respect to the randomness induced by the bounded Laplace mechanism. Again
using~$D = [0, n]$, we have
\begin{align*}
E[g(\tilde{\lambda}_{2})] \!&=\!\! \int_{-\infty}^{\infty} \!\!\!g(x)f_{W_{\lambda_2}}(x)dx \!=\!\! \int_{D} \!g(x)\frac{1}{C_{\lambda_2}(b)}\frac{1}{2b}e^{-\frac{|x-\lambda_{2}|}{b}}dx \\
 &= \frac{1}{C_{\lambda_2}(b)}\frac{1}{2b}\int_{0}^{n}|e^{-xt}-e^{-\lambda_{2}t}|e^{-\frac{|x-\lambda_{2}|}{b}}dx.
\end{align*}
Then, note that
\[
|e^{-xt}-e^{-\lambda_{2}t}|e^{-\frac{|x-\lambda_{2}|}{b}}=\begin{cases}
\left(e^{-xt}-e^{-\lambda_{2}t}\right)e^{\frac{x-\lambda_{2}}{b}} & x\leq\lambda_{2}\\
\left(e^{-\lambda_{2}t}-e^{-xt}\right)e^{\frac{\lambda_{2}-x}{b}} & x>\lambda_{2}
\end{cases}.
\]
Then we eliminate the absolute value to find
\begin{align}
E[g(\tilde{\lambda}_{2})] &=\frac{1}{C_{\lambda_2}(b)}\frac{1}{2b}\int_{0}^{\lambda_{2}}\left(e^{-xt}-e^{-\lambda_{2}t}\right)e^{\frac{x-\lambda_{2}}{b}}dx\\&+\frac{1}{C_{\lambda_2}(b)}\frac{1}{2b}\int_{\lambda_{2}}^{n}\left(e^{-\lambda_{2}t}-e^{-xt}\right)e^{\frac{\lambda_{2}-x}{b}}dx,
\end{align}
and thus
    $E[g(\tilde{\lambda}_{2})]=\frac{1}{C_{\lambda_2}(b)}\frac{1}{2}\Big(\rho_1(t)+\rho_2(t)-\rho_3(t)\Big)$.
Since $g(\tilde{\lambda}_{2})$ is a non-negative random variable,
we can use Markov's inequality to arrive at the theorem statement. 
It can be shown that $\rho_1(t),\ \rho_2(t),\ \rho_3(t)\geq 0$ for all $t.$
\subsection{Proof of Theorem~\ref{thm:times}}
\label{sec:timebound}
To derive a sufficient condition, we fix~$a$ and
upper bound the probability~$\frac{1}{C_{\lambda_2}(b)}\frac{1}{2a}\left(\rho_1(t)+\rho_2(t)-\rho_3(t)\right)$ 
from Theorem~\ref{thm:consensus}. 
Then we find times~$t$ for which this upper bound is bounded above by~$\eta$. 

First, since $\rho_3(t)\geq 0$ for all t,
\begin{align*}
   & \frac{1}{2aC_{\lambda_2}(b)}\left(\rho_1(t)+\rho_2(t)-\rho_3(t)\right) \leq \frac{\rho_1(t)+\rho_2(t)}{2aC_{\lambda_2}(b)}\\& =\frac{e^{-\lambda_{2}t}}{2aC_{\lambda_2}(b)}\left[\left(e^{-\frac{\lambda_{2}}{b}}-e^{\frac{\lambda_{2}-n}{b}}\right)+\frac{e^{\lambda_{2}t}e^{-\frac{\lambda_{2}}{b}}-1}{bt-1}\right].
\end{align*}
By upper bounding again, we can eliminate the $-1$ and $e^{-\frac{\lambda_{2}}{b}}$ in the second to last term giving
\begin{align*}
    &\frac{e^{-\lambda_{2}t}}{2aC_{\lambda_2}(b)}\left[\left(e^{-\frac{\lambda_{2}}{b}}-e^{\frac{\lambda_{2}-n}{b}}\right)+\frac{e^{\lambda_{2}t}e^{-\frac{\lambda_{2}}{b}}-1}{bt-1}\right]&\\
     &\leq\frac{e^{-\lambda_{2}t}}{2aC_{\lambda_2}(b)}\frac{1}{bt-1}\left[\left(bt-1\right)\left(e^{-\frac{\lambda_{2}}{b}}-e^{\frac{\lambda_{2}-n}{b}}\right)+e^{\lambda_{2}t}\right]
     \\&=\frac{e^{-\lambda_{2}t}}{2aC_{\lambda_2}(b)}\left(e^{-\frac{\lambda_{2}}{b}}-e^{\frac{\lambda_{2}-n}{b}}\right)+\frac{1}{2aC_{\lambda_2}(b)\left(bt-1\right)}.
\end{align*}
We have now found an upper bound on $P\left(|\tilde{r}(t)-r(t)|\geq a\right)$, and we analyze times for which
\begin{equation}
\frac{e^{-\lambda_{2}t}}{2aC_{\lambda_2}(b)}\left(e^{-\frac{\lambda_{2}}{b}}-e^{\frac{\lambda_{2}-n}{b}}\right)+\frac{1}{2aC_{\lambda_2}(b)\left(bt-1\right)} \leq\eta,
\end{equation}
which occurs if
\begin{equation} \label{eq:lemma2main}
 e^{-\lambda_{2}t}\left(e^{-\frac{\lambda_{2}}{b}}-e^{\frac{\lambda_{2}-n}{b}}\right)\left(bt-1\right) \leq2aC_{\lambda_2}(b)\eta\left(bt-1\right) -1 . 
\end{equation}

Note that $e^{-\frac{\lambda_{2}}{b}}-e^{\frac{\lambda_{2}-n}{b}} < 0$
when $\lambda_{2}>\frac{n}{2},$ and ${e^{-\frac{\lambda_{2}}{b}}-e^{\frac{\lambda_{2}-n}{b}} \geq 0}$
when $\lambda_{2}\leq\frac{n}{2}.$ We first analyze the case where
$\lambda_{2}\leq\frac{n}{2}$. 
Returning to~\eqref{eq:lemma2main}, 
we find a sufficient condition by using~$bt - 1 \leq bt$ 
and the non-negativity of the parenthetical term. 
Doing so and expanding gives
\begin{equation} \label{eq:lemma2firsthalf}
\left(e^{-\frac{\lambda_{2}}{b}}-e^{\frac{\lambda_{2}-n}{b}}\right)bte^{-\lambda_{2}t}  \leq-2aC_{\lambda_2}(b)\eta-1+2aC_{\lambda_2}(b)\eta bt.
\end{equation}

Next, we maximize the left-hand side over~$t$. 
By setting its time derivative equal to zero, we find that the maximum is attained at the time~$t^*$ satisfying
\begin{equation}
e^{-\lambda_{2}t^{*}}-\lambda_{2}t^{*}e^{-\lambda_{2}t^{*}}=0,
\end{equation}
which implies that~$t^{*}=\frac{1}{\lambda_{2}}.$ Thus, 
\begin{equation}
\sup_{t}\left\{ \left(e^{-\frac{\lambda_{2}}{b}}-e^{\frac{\lambda_{2}-n}{b}}\right)bte^{-\lambda_{2}t}\right\} =\left(e^{-\frac{\lambda_{2}}{b}}-e^{\frac{\lambda_{2}-n}{b}}\right)\frac{b}{\lambda_{2}e}.
\end{equation}

Then, since the right hand side of~\eqref{eq:lemma2firsthalf} grows linearly with time, if we find the time where the right hand side is equal to the maximum
of the left hand side, then the inequality will hold for all $t$
larger than that. Plugging the maximum of the left-hand side into~\eqref{eq:lemma2firsthalf} gives
\begin{equation*}
\left(e^{-\frac{\lambda_{2}}{b}}-e^{\frac{\lambda_{2}-n}{b}}\right)\frac{b}{\lambda_{2}e} =-2aC_{\lambda_2}(b)\eta-1+2aC_{\lambda_2}(b)\eta bt,
\end{equation*}and solving for $t$ gives
\begin{equation*}
t =\frac{\left(e^{-\frac{\lambda_{2}}{b}}-e^{\frac{\lambda_{2}-n}{b}}\right)\frac{b}{\lambda_{2}e}+2aC_{\lambda_2}(b)\eta+1}{2aC_{\lambda_2}(b)\eta b}. 
\end{equation*}
Thus when $\lambda_{2}\leq\frac{n}{2},$ we have~$P\left(|\tilde{r}(t)-r(t)|\geq a\right)\leq\eta$
for $t\geq\frac{\left(e^{-\frac{\lambda_{2}}{b}}-e^{\frac{\lambda_{2}-n}{b}}\right)\frac{b}{\lambda_{2}e}+2aC_{\lambda_2}(b)\eta+1}{2aC_{\lambda_2}(b)\eta b}$.

Now we analyze the case where $\lambda_{2}>\frac{n}{2}$. This causes~$e^{-\frac{\lambda_{2}}{b}}-e^{\frac{\lambda_{2}-n}{b}} < 0$, and
the left-hand side of~\eqref{eq:lemma2main} is negative for all~$t>0$. 
Therefore, we can ensure that~\eqref{eq:lemma2main} holds when the right-hand
side is positive, i.e., when
\begin{equation}
-2aC_{\lambda_2}(b)\eta-1+2aC_{\lambda_2}(b)\eta bt \geq 0. 
\end{equation}
Solving gives~$t \geq\frac{2aC_{\lambda_2}(b)\eta+1}{2aC_{\lambda_2}(b)\eta b}.$

\subsection{Proof of Theorem \ref{thm:expectation_bounds_d_rho}}\label{apdx:thm4}
Since both lower bounds are convex functions with respect to $\lambda_2>0$, we can use Jensen's inequality and we have
    \begin{align}
        &E[\tilde{d}] \geq E[\underline{d}(\tilde{\lambda}_2)] = E\left[\frac{4}{n\tilde{\lambda}_2}\right] \geq \frac{4}{nE[\tilde{\lambda}_2]}\\
        &\begin{multlined}E[\tilde{\rho}] \geq E[\underline{\rho}(\tilde{\lambda_2})] = E\left[\frac{2}{(n-1)\tilde{\lambda}_2} + \frac{n-2}{2(n-1)}\right] \\
        \geq \frac{2}{(n-1)E[\tilde{\lambda}_2]} + \frac{n-2}{2(n-1)}.
        \end{multlined}
    \end{align}
    
    The value of $E[\tilde{\lambda}_2]$ can be computed as
    \begin{align*}
        &E[\tilde{\lambda}_2] = \frac{1}{C_{\lambda_2}(b)}\frac{1}{2b}\int_0^n x e^{-\frac{|x-\lambda_2|}{b}}dx \\
        &=\frac{1}{C_{\lambda_2}(b)}\frac{1}{2b}\left(\int_0^{\lambda_2}xe^{-\frac{\lambda_2-x}{b}}dx + \int_{\lambda_2}^n xe^{-\frac{x-\lambda_2}{b}}dx \right)\\
        &= \frac{1}{2C_{\lambda_2}(b)}\left(2\lambda_2+be^{-\frac{\lambda_2}{b}}-be^{-\frac{n-\lambda_2}{b}}-ne^{-\frac{n-\lambda_2}{b}}\right).
    \end{align*}
    
    We next compute the expectation term $E\left[\frac{1}{\sqrt{\tilde{\lambda}_2}}\right]$ as
    
    \begin{align*}
        &E\left[\frac{1}{\sqrt{\tilde{\lambda}_2}}\right] = \frac{1}{C_{\lambda_2}(b)}\frac{1}{2b}\int_0^n \frac{1}{\sqrt{x}} e^{-\frac{|x-\lambda_2|}{b}}dx \\
        &=\frac{1}{C_{\lambda_2}(b)}\frac{1}{2b}\left(\int_0^{\lambda_2}\frac{1}{\sqrt{x}} e^{-\frac{\lambda_2-x}{b}}dx + \int_{\lambda_2}^n\frac{1}{\sqrt{x}} e^{-\frac{x-\lambda_2}{b}}dx \right)\\
        &= \frac{1}{C_{\lambda_2}(b)}\frac{1}{2b} \left(\sqrt{\pi}\sqrt{b}e^{-\frac{\lambda_2}{b}}\left(\textrm{erfi}\left(\sqrt{\frac{\lambda_2}{b}}\right)\right)\right.\\
        & \quad\quad\left.+\sqrt{b}e^{\frac{\lambda_2}{b}}\left(\Gamma\left(\frac{1}{2},\frac{n}{b}\right)-\Gamma\left(\frac{1}{2},\frac{\lambda_2}{b}\right)\right)\right).
    \end{align*}
    
    Then we can find the desired upper bounds by applying the linearity of expectation.\hfill $\blacksquare$


\end{document}